\documentclass[11pt]{article}
\pdfoutput=1

\usepackage[usenames,dvipsnames]{xcolor}
\usepackage[a4paper]{geometry}
\usepackage[english]{babel}
\usepackage{cite,enumerate,enumitem,booktabs,float,graphicx}

\usepackage[T1]{fontenc}
\usepackage[utf8]{inputenc}
\usepackage{lmodern}

\makeatletter
\g@addto@macro\bfseries{\boldmath}
\makeatother

\usepackage{bm,amsmath,amssymb,tensor,mathtools}
\numberwithin{equation}{section}

\usepackage{mathrsfs}		
\usepackage[scr=rsfs]{mathalfa}

\usepackage{feynmp-auto}

\usepackage{caption}
\usepackage{subcaption}

\usepackage[pdftex]{hyperref}
\definecolor{dark-blue}{rgb}{0.15,0.15,0.4}
\hypersetup{
  colorlinks,
  linkcolor={dark-blue},
  citecolor={blue}
}

\newcommand{\beq}{\begin{equation}}
\newcommand{\eeq}{\end{equation}}
\DeclareMathOperator{\atanh}{atanh}

\DeclareMathOperator{\re}{Re}
\newcommand{\SO}{\mathrm{SO}}

\renewcommand{\SS}{\mathcal{S}}

\usepackage{authblk}

\title{\textbf{%
	Scattering amplitudes for \\cylindrical gravitational waves
  }
}
\date{}

\author[]{Robert Penna\footnote{pennar@sunypoly.edu}}
\affil[]{%
  Department of Mathematics and Physics,
  \protect\\
  SUNY Polytechnic Institute,
  Utica, NY 13502 USA
}

\begin{document}
\unitlength = 0.8mm 
\maketitle
\thispagestyle{empty}

\begin{abstract}

Cylindrical gravitational waves are interesting because they enjoy an infinite dimensional symmetry called Geroch symmetry.  
In this paper, we compute the 2-particle tree-level $\SS$-matrix for cylindrical gravitational waves.  
The model we use is a dimensional reduction of general relativity to two spacetime dimensions.  
The reduced theory is a nonlinear sigma model.  
We discuss an $\SO(2)$ symmetry of the amplitudes which is a special case of Geroch symmetry.

\end{abstract}

\newpage


\section{Introduction}

One of the reasons cylindrical gravitational waves are interesting is that they enjoy a distinctive infinite dimensional symmetry called Geroch symmetry \cite{Geroch:1970nt,Geroch:1972yt,Julia:1981wc,Breitenlohner:1986um,Nicolai:1991tt,Schwarz:1995af,Schwarz:1995td,Grumiller:2000wt,Lu:2007jc,Lu:2007zv}.  Geroch symmetry renders the classical theory exactly solvable.  An intriguing open question is whether it does the same for the quantum theory.  If cylindrical gravitational waves have an exactly solvable $\SS$-matrix, then that would probably be a nice toy model for quantum gravity.  

In this paper, we take some modest steps in this direction.  We compute the 2-particle tree-level $\SS$-matrix for cylindrical gravitational wave scattering.  The model we use is a dimensional reduction of general relativity to two spacetime dimensions.  The reduced theory is a nonlinear sigma model.  We quantize the sigma model and study its scattering amplitudes\footnote{A number of authors have studied the quantization of cylindrical gravitational waves as a way to learn general lessons about quantum gravity\cite{Kuchar:1971xm,Ashtekar:1996bb,Ashtekar:1997zf,Niedermaier:1999bh,Niedermaier:2000ud,Niedermaier:2002eq,Niedermaier:2003fz,BarberoG:2003ffm,BarberoG:2003njp,BarberoG:2004kdm,BarberoG:2008fcp,BarberoG:2010oga}.  Our amplitude calculations appear to be new.}.  This sigma model is known to describe classical cylindrical gravitational waves perfectly.  Since tree-level physics is essentially classical physics, we should be able to trust the sigma model at tree level, and that is all we will consider in this work.

The sigma model has two scalar fields, $X^1$ and $X^2$, which correspond to the two physical polarizations of the graviton.  
The theory has an $\SO(2)$ symmetry which in the free theory just rotates the vector $X^\mu$ ($\mu=1,2$).  
This symmetry is a simple example of a Geroch transformation.  
We will show below that by combining this symmetry with crossing symmetry we can reduce the 2-particle tree-level $\SS$-matrix to a single function. 
An interesting problem for the future is to understand the full set of constraints imposed by Geroch symmetry on the $\SS$-matrix.  

In this paper we compute the amplitudes directly using Feynman diagrams. 
The $\SS$-matrix for ordinary gravitons has a famous soft pole.  
The cylindrical gravitational wave $\SS$-matrix we compute herein has a soft zero. 
In a forthcoming work, we will use Geroch symmetry to give a symmetry explanation for this soft zero \cite{penna2024}.  

The present work leaves a number of open questions for the future but the most pressing is to extend our calculations to one loop and understand if Geroch symmetry persists or if it is spoiled by anomalies.

\section{Sigma model}

The two-dimensional spacetime metric is
\beq
ds^2 = -dt^2 + dr^2 \,.
\eeq
Space is a half-line ($r>0$).  
Future and past infinity, $\Sigma^\pm$, are the lines at $t=\pm\infty$ with coordinate $r$.  
The Lagrangian of the sigma model is \cite{Breitenlohner:1986um,Nicolai:1991tt,Schwarz:1995af,Schwarz:1995td,Grumiller:2000wt,Lu:2007jc,Lu:2007zv}
\beq\label{eq:L}
L	=	- \frac{r}{2} (\partial_a X^1) (\partial^a X^1) 
		- \frac{r}{2} e^{2X^1} (\partial_a X^2) (\partial^a X^2) \,.
\eeq
For notational simplicity we have set the coupling $g=1$.  
Here $X^1=X^1(t,r)$ and $X^2=X^2(t,r)$ are scalar fields.  
They encode the two physical polarizations of the graviton.  
Eq. \eqref{eq:L} has a global symmetry
\beq\label{eq:so2}
\delta X^1 = -X^2 \,, \qquad
\delta X^2 =  \frac{1}{2} \left[ 1 - e^{-2 X^1} + ( X^2 )^2 \right]   .
\eeq
In the free theory, this is just an $\SO(2)$ rotation of the vector $X^\mu$.   
Eq. \eqref{eq:so2} is a simple example of a Geroch transformation.  
The full set of Geroch transformations is infinite dimensional.  Most of the Geroch symmetries are hidden symmetries in the sense that they involve nonlocal transformations of the fields.  

The sigma model reduces to a free theory on $\Sigma^\pm$.  In the free theory, the Noether charge for eq. \eqref{eq:so2} is
\beq\label{eq:Q}
Q^\pm = \int_{\Sigma^\pm} dr \, r \left[ X^2 \partial_t X^1 - X^1 \partial_t X^2 \right] .
\eeq
The mode expansion of the free field is (suppressing the $\pm$)
\beq\label{eq:modes}
X(t,r)
	= \frac{1}{\sqrt{2}} \int_0^\infty dk 
		\left[ a(k) e^{-ikt} + a(k)^\dag e^{ikt} \right] J_0(k r) \,.
\eeq
This is an expansion in energy eigenmodes\footnote{There is another family of energy eigenmodes ($e^{\pm ikt} Y_0(kr)$) but they are singular at $r=0$ so we do not include them.}.
The creation and annihilation operators, $a^\dag$ and $a$, obey
\beq
\left[ a^\mu(k) , a^\nu(k')^\dag \right] = \delta^{\mu\nu} \delta(k - k') \,.
\eeq
Inserting eq. \eqref{eq:modes} into \eqref{eq:Q} and doing the $r$ integral gives
\beq
Q = i \int dk \left[ a^1(k)^\dag a^2(k) - a^2(k)^\dag a^1(k) \right] .
\eeq
Thanks to this symmetry, the 2-particle tree-level $\SS$-matrix takes the form
\beq\label{eq:so2eq}
\tensor[_{\mu\rho}]{\SS}{_{\nu\gamma}} = \vcenter{ \hbox{ \includegraphics[width=3cm]{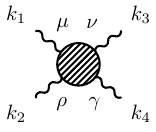} } }
	=	\delta_{\mu\rho} \delta_{\nu\gamma} \sigma_1 
		+ \delta_{\mu\nu} \delta_{\rho\gamma} \sigma_2 
		+ \delta_{\mu\gamma} \delta_{\nu\rho} \sigma_3 \,.
\eeq
So there are just three functions to consider: $\sigma_1$, $\sigma_2$, and $\sigma_3$.  And in fact the $\sigma_i$'s are all related by crossing relations, so there is really only one independent function.  Indeed, $\sigma_2(k_1,\dots, k_4)$ and $\sigma_3(k_1,\dots, k_4)$ are simply related by interchanging $k_3$ and $k_4$.  And $\sigma_1(k_1, \dots, k_4)$ and $\sigma_2(k_1, \dots, k_4)$ are simply related by interchanging $k_2$ and $k_3$ and putting in an overall minus sign.  
In the next section, we will pick one of these functions ($\sigma_1$) and compute it using Feynman diagrams.

\section{Feynman diagrams}

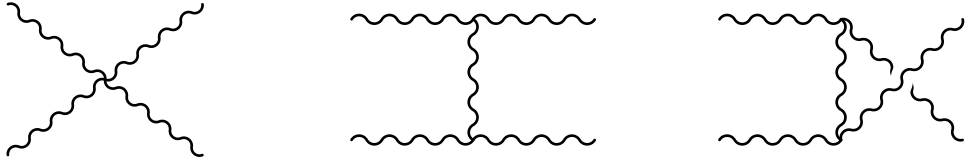
\begin{figure}
\vspace{1cm}
\centering
\begin{fmffile}{sigma1c}
\begin{fmfgraph*}(40,25)
\fmfleft{i1,i2}
\fmfright{o1,o2}
\fmf{photon}{i1,o2}
\fmf{photon}{i2,o1}
\end{fmfgraph*}
\end{fmffile}
\hspace{1cm}
\begin{fmffile}{sigma1t}
\begin{fmfgraph*}(40,25)
\fmftop{i1,o1}
\fmfbottom{i2,o2}
\fmf{photon}{i1,v1,o1}
\fmf{photon,tension=0}{v1,v2}
\fmf{photon}{i2,v2,o2}
\end{fmfgraph*}
\end{fmffile}
\hspace{1cm}
\begin{fmffile}{sigma1u}
\begin{fmfgraph*}(40,25)
\fmftop{i1,o1}
\fmfbottom{i2,o2}
\fmf{photon}{i1,v1}
\fmf{photon,tension=0}{v1,v2}
\fmf{photon}{i2,v2}
\fmf{phantom}{v1,o1}
\fmf{phantom}{v2,o2}
\fmf{photon,tension=0}{v1,o2}
\fmf{photon,tension=0,rubout=8}{v2,o1}
\end{fmfgraph*}
\end{fmffile}
\caption{Tree-level contributions to $\sigma_1$.}
\label{fig:sigma1}
\end{figure}

The goal of this section is to compute $\sigma_1$ using Feynman diagrams.  There are three diagrams to consider (Fig. \ref{fig:sigma1}).  Actually, the second and third diagrams are basically the same (just interchange $k_3$ and $k_4$), so we really only need to consider the first two diagrams.  The first diagram comes from the quartic interaction term \eqref{eq:L}:
\beq
- r (X^1)^2 (\partial_a X^2) (\partial^a X^2) \,.
\eeq
The amplitude is
\beq
	- 4i \int d^2 x \, r \, e^{-i (k_1 + k_2) t } J_0(k_1 r) J_0(k_2 r) 
		\partial_a [e^{i k_3 t} J_0(k_3 r)] \partial_a [e^{i k_4 t} J_0(k_4 r)] \,.
\eeq
Expanding the derivatives and doing the $t$ integral gives
\beq
- 8\pi i k_3 k_4 \delta_{k_3 + k_4 - k_1 - k_2}
	\int dr \, r \, J_0(k_1 r) J_0(k_2 r) [ J_0(k_3 r) J_0(k_4 r) + J_1(k_3 r) J_1(k_4 r) ] \,.
\eeq
To simplify this further, we use the identity
\beq\label{eq:Jidentity}
J_0(a r) J_0(b r)
	= \frac{1}{2\pi} \int_{|a-b|}^{a+b} dc \frac{c J_0(c r)}{\Delta(a,b,c)} \,.
\eeq
$\Delta(a,b,c)$ is the area of the triangle\footnote{A convenient formula is $\Delta = \frac{1}{4} \sqrt{ (a + b)^2 - c^2 } \sqrt{ c^2 - ( a - b )^2 }$.} with side lengths $a$, $b$, and $c$.  
Using eq. \eqref{eq:Jidentity} we obtain
\begin{align}
- 4 i k_3 k_4 \delta_{k_3 + k_4 - k_1 - k_2}
	&\int_{|k_1 - k_2|}^{k_1 + k_2} \frac{dk \, k}{\Delta(k_1, k_2, k)} 			\notag \\
	&\cdot \int dr \, r \, J_0(k r) [ J_0(k_3 r) J_0(k_4 r) + J_1(k_3 r) J_1(k_4 r) ] \,.
\end{align}
The $r$ integral on the second line is zero unless it is possible to form a triangle with side lengths $k$, $k_3$, and $k_4$, in which case we have
\begin{align}
\int dr \, r  J_0(a r) J_0(b r) J_0(c r) &= \frac{1}{2\pi \Delta(a,b,c)} \,,				\label{eq:sonine1}	\\
\int dr \, r  J_0(a r) J_1(b r) J_1(c r) &= \frac{b^2 + c^2 - a^2}{4 \pi b c \Delta(a,b,c)}	\label{eq:sonine2}	 \,.
\end{align}
We thus obtain
\begin{align}
- \frac{16 i}{\pi} \delta_{k_3 + k_4 - k_1 - k_2}
	\int_{{\rm max}(|k_1 - k_2|,|k_3 - k_4|)}^{k_1 + k_2} 
	\frac{k \, dk}{\sqrt{k^2 - (k_1 - k_2)^2} \sqrt{k^2 - (k_3 - k_4)^2}} \,.
\end{align}
Finally, we perform the $k$ integral to obtain
\begin{align}
- \frac{16 i}{\pi} \re \atanh \sqrt{\frac{k_1 k_2}{k_3 k_4}} \delta_{k_3 + k_4 - k_1 - k_2}  \,.
\end{align}
This is the amplitude associated to the first diagram in Fig. \ref{fig:sigma1}.

Now turn to the second diagram in Fig. \ref{fig:sigma1}.  
This diagram is slightly more complicated because it has an internal line.  
To put in the internal line, we need the Feynman propagator:
\beq\label{eq:feynman}
\langle X^\mu_x X^\nu_{x'} \rangle_0
	=  \delta^{\mu\nu}\frac{i}{2\pi}  \int d\omega \int dk \,
		\frac{k}{\omega^2 - k^2 + i \epsilon} e^{ -i\omega (t-t') } J_0(k r) J_0(k r')  \,.
\eeq
The vertices in this diagram come from the cubic interaction \eqref{eq:L}:
\beq
-r X^1 (\partial_a X^2) (\partial^a X^2) \,.
\eeq
Putting everything together, the amplitude is
\begin{align}
- \frac{2i}{\pi}  &\int d\omega \int dk \, \frac{k}{\omega^2 - k^2 + i \epsilon}  		\notag \\ 
&	\cdot \int d^2 x \, r \, e^{-ik_1 t} J_0(k_1 r) \partial_a [e^{i k_3 t} J_0(k_3 r)] \partial^a [e^{ -i\omega t } J_0(k r) ]	\notag \\
&	\cdot \int d^2 x' \, r'  \, e^{-ik_2 t'} J_0(k_2 r') \partial_{a'} [e^{i k_4 t'} J_0(k_4 r')] \partial^{a'}[ e^{ i\omega t' } J_0(k r')] \,.
\end{align}
Expanding the derivatives and doing the $t$ and $\omega$ integrals, we obtain
\begin{align}
8\pi i k_3 k_4 & \delta_{k_3 + k_4 - k_1 - k_2}  \int dk \, \frac{k}{k^2 - (k_3 - k_1)^2}  		\notag \\ 
&	\cdot \int d^2 x \, r \, J_0(k_1 r) [ - (k_3 - k_1) J_0(k_3 r) J_0(k r) + k J_1(k_3 r) J_1(k r)  ]  \notag \\
&	\cdot \int d^2 x' \, r'  \, J_0(k_2 r')[ (k_2 - k_4) J_0(k_4 r') J_0(k r') + k J_1(k_4 r')J_1(k r') ] \,.
\end{align}
Next, we do the $r$ integrals using eqs. \eqref{eq:sonine1} and \eqref{eq:sonine2} to get
\begin{align}
\frac{8 i}{\pi} & \delta_{k_3 + k_4 - k_1 - k_2}  
	\int_{|k_1 - k_3|}^{{\rm min}(k_1 + k_3, k_2 + k_4)} dk \, \frac{k}{\sqrt{(k_1 + k_3)^2 - k^2} \sqrt{(k_2 + k_4)^2 - k^2}} \,.
\end{align}
This integral deserves a comment.  It might seem strange that we are integrating over the ``momentum,'' $k$, of the internal line in a tree diagram.  
But our Lagrangian \eqref{eq:L} does not have $r$-translation invariance, so $k$ is not conserved at vertices.   Mathematically, this manifests in the fact that the triple Bessel integrals \eqref{eq:sonine1} and \eqref{eq:sonine2} are not $\delta$-functions\footnote{Our Lagrangian does have time-translation invariance and this is why we do get an overall energy-conserving $\delta$-function, $\delta_{k_3 + k_4 - k_1 - k_2}$.}.  Performing the $k$ integral gives
\beq
\frac{8 i}{\pi} \re\atanh \sqrt{\frac{ k_1 k_3 }{ k_2 k_4 }} \delta_{k_3 + k_4 - k_1 - k_2}  \,.
\eeq
This is the amplitude associated to the second diagram in Fig. \ref{fig:sigma1}.  The third diagram is almost the same, we just exchange $k_3$ and $k_4$.  
Summing all three diagrams gives 
\beq
\sigma_1 = - \frac{8 i}{\pi} \re\atanh \sqrt{\frac{ k_1 k_2 }{ k_3 k_4 }} \delta_{k_3 + k_4 - k_1 - k_2}  \,.	\label{eq:sigma1}
\eeq
Now using the crossing relations discussed in the previous section, we obtain
\begin{align}
\sigma_2 =	\frac{8 i}{\pi} \re\atanh \sqrt{\frac{ k_1 k_3 }{ k_2 k_4 }} 
			\delta_{k_3 + k_4 - k_1 - k_2}  \,,		\label{eq:sigma2} \\
\sigma_3 =	\frac{8 i}{\pi} \re\atanh \sqrt{\frac{ k_1 k_4 }{ k_2 k_3 }} 
			\delta_{k_3 + k_4 - k_1 - k_2}  \,.		\label{eq:sigma3}
\end{align}
A couple consistency checks are in order.  It is clear from the Lagrangian \eqref{eq:L} that there are no possible tree diagrams for $\tensor[_{11}]{\SS}{_{11}} = \sigma_1 + \sigma_2 + \sigma_3$, so the sum on the right hand side should vanish.  And indeed it is not hard to sum eqs. \eqref{eq:sigma1}--\eqref{eq:sigma3} and check that $\sigma_1 + \sigma_2 + \sigma_3 = 0$.  
Furthermore, if we look at eq. \eqref{eq:so2eq} we can observe 
$\sigma_1 + \sigma_2 + \sigma_3 = \tensor[_{22}]{\SS}{_{22}}$.  The amplitude on the right hand side does receive contributions from three diagrams but we have computed these diagrams (we omit the details for brevity) and indeed they sum to zero.  

The amplitudes we have just computed have the curious feature that they vanish in the soft limit,  $k_i \rightarrow 0$.  In a forthcoming work, we will show how the Geroch group provides a symmetry explanation for this soft zero \cite{penna2024}.

\subsection*{Acknowledgements}

I am grateful to Daniel Grumiller, Bart Horn, and Max Niedermaier for comments on an early version of this work.

\bibliographystyle{JHEP}
\bibliography{biblio}

\end{document}